\documentclass[aps, a4paper, twocolumn, nofootinbib, showpacs, amsfonts]{revtex4}

\usepackage[utf8]{inputenc}
\usepackage{ dsfont }
\usepackage{amssymb}

\usepackage{graphicx}
\usepackage{amsmath}
\usepackage{bm}
\usepackage{xcolor}

\usepackage{tikz}
\usepackage{mathrsfs}
\usetikzlibrary{arrows}
\usetikzlibrary{calc,decorations.pathmorphing,shapes}
\usepackage{fancyhdr}
\usepackage[utf8]{inputenc}
\newcommand{\bea}{\begin{eqnarray}}
\newcommand{\eea}{\end{eqnarray}}
\newcommand{\be}{\begin{equation}}
\newcommand{\ee}{\end{equation}}

\definecolor{rvwvcq}{rgb}{0.08235294117647059,0.396078431372549,0.7529411764705882}
\definecolor{wrwrwr}{rgb}{0.3803921568627451,0.3803921568627451,0.3803921568627451}

\begin{document}

\title{Dust collapse in asymptotic safety: a path to regular black holes}

\author{Alfio Bonanno}
\email{alfio.bonanno@inaf.it}
\affiliation{INAF, Osservatorio Astrofisico di Catania, via S.Sofia 78, I-95123 Catania, Italy}
\affiliation{INFN, Sezione di Catania, via S.Sofia 64, I-95123 Catania, Italy}

\author{Daniele Malafarina}
\email{daniele.malafarina@nu.edu.kz}
\affiliation{Department of Physics, Nazarbayev University, Kabanbay Batyr 53, 010000 Astana, Kazakhstan}

\author{Antonio Panassiti}
\email{antonio.panassiti@phd.unict.it}
\affiliation{INAF, Osservatorio Astrofisico di Catania, via S.Sofia 78, I-95123 Catania, Italy}
\affiliation{INFN, Sezione di Catania, via S.Sofia 64, I-95123 Catania, Italy}
\affiliation{Dipartimento di Fisica e Astronomia “Ettore Majorana”, Universit\`a di Catania, Via S.
Sofia 64, 95123, Catania, Italy}

\begin{abstract}
Regular black hole spacetimes are obtained from an effective Lagrangian for Quantum Einstein Gravity. The interior matter is modeled as a dust fluid, which interacts with the geometry through a multiplicative coupling function denoted as $\chi$. 
The specific functional form of $\chi$ is deduced from Asymptotically Safe gravity, under the key assumption that the Reuter fixed point remains minimally affected by the presence of matter. As a consequence the gravitational coupling vanishes at high energies. 
The static exterior geometry of the black hole is entirely determined by the junction conditions at the  boundary surface. 
Consequently, the resulting global spacetime geometry remains devoid of singularities at all times.
{This outcome offers a new perspective on how regular black holes are formed through gravitational collapse.}

\end{abstract}

\maketitle

In the realm of general relativity, black holes (BH) are fascinating objects characterized by spacetime singularities concealed within an event horizon \cite{Hawking:1970zqf}.  
The occurrence of singularities makes a compelling case for the study of models beyond general relativity, where spacetime remains geodesically complete.  One prominent approach to achieve this involves replacing the singularity with a regular patch of de Sitter space \cite{Frolov:1989pf}, an old concept that has garnered renewed attention in recent times. Much of the existing research in the literature derives regular black hole geometries through modifications of the static Misner-Sharp mass with the aim of rapid convergence to zero at small distances. This is the case for static regular BH metrics like the Poisson-Israel model \cite{Poisson:1988wc}, the Asymptotically Safe (AS) model \cite{Bonanno:2000ep}, the Dymnikova regular black hole model \cite{Dymnikova:1992ux,Platania:2019kyx}, or the Hayward metric \cite{Hayward:2005gi}, to name a few (see \cite{Bambi:2023try} for an extended review). 

Yet, attempts to derive these solutions from a physically plausible Lagrangian, such as in the instance of regular black holes in non-linear electrodynamics
\cite{Ayon-Beato:1998hmi}, have faced numerous challenges \cite{Dymnikova:2004zc}. It is reasonable to acknowledge that a unanimous consensus on incorporating regular black hole solutions within a broader gravitational theory remains elusive \cite{Giacchini:2021pmr,Knorr:2022kqp}.

An alternative route to obtaining  modifications of the classical BH solutions is by matching a non-singular homogeneous interior model, describing collapsing matter, to a static exterior black hole solution. This method has received considerable attention in recent years. The minisuperspace approximation in Loop Quantum Cosmology leads to an effective Friedman equation with repulsive (i.e., negative) gravity at high densities, resulting in a bouncing interior model \cite{Hossenfelder:2009fc,Bojowald:2005qw,Bambi:2013caa, Lewandowski:2022zce}.  Conversely, assuming a Renormalization Group (RG) improved regular exterior allows for obtaining a singularity-free collapsing dust model as the interior solution \cite{Torres:2014gta}. 
Other methods obtain similar results relying on different approaches to quantization (see for example \cite{Husain:2021ojz, Calmet:2021stu, Giesel:2021dug}). 
However, similarly to the above mentioned static solutions, deducing these models from an effective Lagrangian  remains challenging.  On the other hand  numerical simulations of the formation of regular BH  are often limited to a 2D dilaton gravity model \cite{Kunstatter:2015vxa,Biasi:2022ktq} and the generalization to a 4D model is problematic unless a physically motivated  Lagrangian formulation is found. 

Within this study, we present a resolution to this quandary by extending an initial idea  by Markov and Mukhanov \cite{Markov:1985py}. Our approach involves formulating gravity's antiscreening behavior in ultraplanckian energy domains \cite{Polyakov:1993tp} through the inclusion of a multiplicative coupling with the matter Lagrangian. The structure of this coupling is guided by the Reuter fixed point of AS gravity \cite{Reuter:1996cp}. Remarkably both the energy-momentum and the effective energy-momentum tensors are conserved in this theory, in contrast to what happens in most approaches mentioned above.
Notably, under conditions of low energy, our model seamlessly recovers the equations of standard general relativity. Also, over extensive distances, the solution for black holes bears resemblance to the Schwarzschild solution.

To be more specific, let us consider a matter fluid with a proper density $\epsilon$, characterized by a 4-velocity $u^\mu$ such that $u_\mu u^\mu = -1$, and a rest-mass density $\rho$. The mass continuity equation is expressed as $(\rho u^\mu)_{; \mu} = 0$, and for a non-dissipative fluid, the relative variations of density are related as 
$\delta\rho/\rho=\delta\epsilon/(p(\epsilon)+\epsilon)$.

Following the approach in \cite{Markov:1985py}, we introduce the action for our system as follows:
\begin{equation}
\label{effS}
S = \frac{1}{16 \pi G_N} \int d^4 x \sqrt{-g} \left[R + 2  \chi(\epsilon)  \mathcal{L}\right] .
\end{equation}
Here, $\mathcal{L} = -\epsilon$ represents the matter Lagrangian, and the function $\chi = \chi(\epsilon)$ serves as a multiplicative gravity-matter coupling with the important property $\chi(\epsilon=0) = 8 \pi G_N$.
The metric variation of the matter part of the Lagrangian yields
\begin{equation}
\frac{1}{\sqrt{-g}}\,\delta
\left(2\,\sqrt{-g}\,\chi\,\epsilon \right) =2\frac{\partial (\chi\epsilon)}{{\partial\epsilon}}\delta\epsilon -\chi\,\epsilon\,g_{\mu\nu}\,\delta g^{\mu\nu}
\ .
\end{equation}
Note that the variation of $\rho$ under a change of the metric is given by \cite{Groen:2007zz}:
\begin{equation}
\delta \rho = \frac{\rho}{2}(g_{\mu\nu} + u_\mu u_\nu )\delta g^{\mu\nu}.
\end{equation}
As a result, the total variation of the action (\ref{effS}) leads to the following field equations:
\begin{equation}
\label{effectiveEQ}
R_{\mu\nu}- \frac{1}{2}g_{\mu\nu}R = \frac{\partial (\chi \epsilon)}{\partial\epsilon}
T_{\mu\nu}+\frac{\partial \chi}{\partial \epsilon} \epsilon^2 g_{\mu\nu} \equiv T_{\mu\nu}^{\rm eff},
\end{equation}
where
\begin{equation}
\label{efeg}
8\pi G(\epsilon)=\frac{\partial (\chi \epsilon)}{\partial \epsilon}, \quad \Lambda(\epsilon)=-\frac{\partial \chi}{\partial \epsilon} \epsilon^2,
\end{equation}
represent the effective Newton constant and cosmological constant, respectively. Here, $T_{\mu\nu}= (\epsilon+p(\epsilon)) u_\mu u_\nu+p g_{\mu\nu}$ is conserved.
For spherically homogeneous collapse the metric functions in the diagonal line element in co-moving coordinates $\{t,r,\theta,\phi\}$ are $g_{00}=-e^{2\nu(r,t)}$, $g_{11}=e^{2\psi(r,t)}$, and $g_{22}=C^2(r,t)$. 
Then we can express the field equations \eqref{effectiveEQ} as follows 
\begin{eqnarray}
\label{G00}
&&\frac{F_{\rm eff}'}{C^2C'} = 8 \pi G(\epsilon)\epsilon  + \Lambda(\epsilon) =\chi(\epsilon) \epsilon \equiv \epsilon_{\rm eff},\\
\label{G11}
&&-\frac{\dot{F}_{\rm eff}}{C^2\dot{C}} = 8 \pi G(\epsilon)p-\Lambda(\epsilon) \equiv p_{\rm eff},
\end{eqnarray}
and
\begin{equation}
\label{G01}
\dot{C}'= \dot{C}\nu'+C'\dot{\psi} .
\end{equation}
Note that $\epsilon_{\rm eff}>0$ always, while $p_{\rm eff}$ can become negative.  Dotted quantities represent derivatives with respect to the comoving time ($t$), while prime denotes derivatives with respect to the comoving radial coordinate ($r$). The function $F_{\rm eff}$ represents the effective Misner-Sharp mass of the system, which can be defined in analogy with the Schwarzschild mass $M$ from $1-F_{\rm eff}/C=g_{\mu\nu}\nabla^\mu C\nabla^\nu C$ \cite{Misner:1964je}, and it is given by:
\begin{equation}
\label{ms}
F_{\rm eff}=C(1-C'^2e^{-2\psi}+\dot{C}^2e^{-2\nu}).
\end{equation}
Additionally, the Bianchi identity takes the form:
\begin{equation}
\nu'=-\frac{p_{\rm eff}'}{\epsilon_{\rm eff}+p_{\rm eff}}.
\end{equation}
In the case of a homogeneous perfect fluid, we have $\epsilon(t)$ and $p(t)$, which in turn imply $\epsilon_{\rm eff}(t)$ and $p_{\rm eff}(t)$. As a consequence of the Bianchi identity, we can set $\nu=0$ and integrate equation \eqref{G01} to obtain $e^{2\psi}=C'^2/(1-Kr^2)$, where $K$ is the integration constant related to the curvature of the 3-space.
The line element can then be written as:
\begin{equation}
ds^2=-dt^2+\frac{C'^2}{1-Kr^2}dr^2+C^2d\Omega^2,
\end{equation}
where $d\Omega^2$ represents the metric on the unit 2-sphere. By rescaling the area-radius function $C$ using the adimensional scale factor $a$ according to $C=ra$,
the energy-momentum conservation equation gives
\begin{equation}
d\epsilon + 3(p(\epsilon)+\epsilon) d \ln a = 0 .
\end{equation}
Following \cite{Malafarina2023} we rescale also the effective Misner-Sharp mass as $F_{\rm eff}=r^3m_{\rm eff}$. We can then rewrite  equations \eqref{G00} and \eqref{G11} as follows:
\begin{equation}
\label{G00-2}
\epsilon_{\rm eff}= \frac{3m_{\rm eff}}{a^3}, \quad \quad
p_{\rm eff}= -\frac{m_{{\rm eff},a}}{a^2},
\end{equation}
where by $X_{,a}$ we indicate derivatives of $X$ with respect to $a$,
and equation \eqref{ms} becomes:
\begin{equation}
m_{\rm eff}=a(\dot{a}^2+K) .
\end{equation}
Here, $m_{\rm eff}(a)=- aV(a)$, where the potential $V(a)$ reads 
\begin{equation}\label{V(a)}
V(a) = -\frac{8 \pi}{3}a^2\int_0^{\epsilon(a)} G(s) ds .
\end{equation}
In principle, if we know $G(\epsilon)$ from a fundamental theory, from (\ref{efeg}) is possible to determine $\chi$ and close the system. 

In this work, we assume that the behavior of $G$ as a function of the energy scale is governed by a renormalization group trajectory close to the ultraviolet (UV) fixed point of the AS program \cite{Weinberg:1980gg, Reuter:1996cp, Eichhorn:2018yfc, Pawlowski:2020qer, Reichert:2020mja, Reuter:2019byg, Percacci:2017fkn, Bonanno:2019ilz} 
and we neglect higher order operators in $R$ in the effective Lagrangian.

As demonstrated in \cite{Falls:2014tra} and \cite{Falls:2017lst}, they are all irrelevant in the language of the renormalization group, therefore their impact on the renormalized trjaectories around the Reuter fixed point can be neglected.
The $R^2$  operator is instead classically marginal, however given its considerably less prominent UV scaling compared to the running of the $R$ operator \cite{Falls:2014tra} we will not consider this term.  On the contrary, the cosmological constant emerges as a relevant operator, but its scaling behavior is inherently encompassed in the Markov-Mukhanov approach through the emergence of an effective cosmological constant in equation (\ref{effectiveEQ}).

 We thus adopt the approximate running of $G$ as proposed in \cite{Bonanno:2021squ} in the limit of Quantum Einstein Gravity
\begin{equation}
\label{trajectory}
G(k) = \frac{G_N}{1+G_Nk^2/g_*},
\end{equation}
where $k$ represents the IR regulator scale, and $g_\ast=570\pi/833$ is the UV fixed point. To connect $k$ to $\epsilon$, we require a prescription.  The analysis in \cite{Bonanno:2021squ} demonstrates that the UV scaling of the physical Newton constant $G_N(q)$, where $q$ is the external momentum, is in essence the same as predicted by the renormalized coupling $G(k)$. The only distinction lies in the crossover scale, a non-universal feature of the flow (see particularly Fig.4 in \cite{Bonanno:2021squ}). 
Hence the identification $k^2\sim q^2 \sim 1/d^2$  where $d$ is a  distance scale 
should  capture the qualitative features of the physical flow 
 \cite{Bonanno:2020bil}.

However, it is essential to acknowledge that the key assumption underlying equation (\ref{trajectory}) is that the presence of matter does not significantly deform the flow and compromise the fixed point \cite{Dona:2013qba}. This assumption is crucial in maintaining the integrity of the renormalization group trajectory and ensuring its consistency in the presence of matter.

\begin{figure}[tt]
\includegraphics[width=0.97\linewidth]{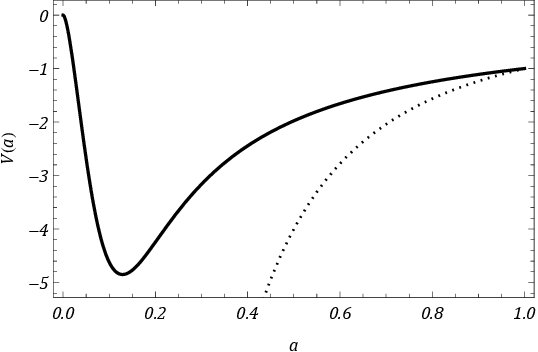}
\caption{The thick line shows the potential  $V(a)$ for dust collapse in the AS model with running $G$ given by equation \eqref{G}. For comparison the OSD model with $8\pi G_N=1$ is shown as the dotted line. The parameters are fixed for illustrative purposes to $m_0=1$ and $\xi=0.001$.} \label{fig:V}
\end{figure}

\begin{figure}[tt]
\includegraphics[width=0.97\linewidth]{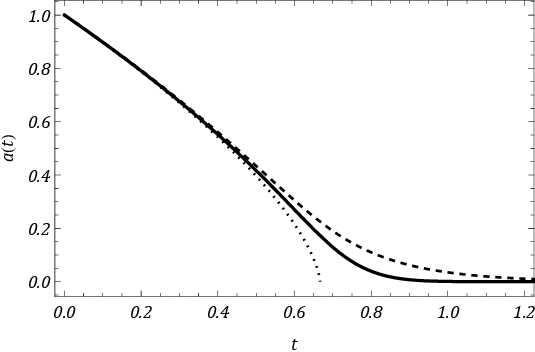}
\caption{The thick line shows the scale factor $a(t)$ for marginally bound ($K=0$) dust collapse in the AS model with running gravitation and cosmological constant solution of equation \eqref{eom}. For comparison the OSD collapse model is shown as the dotted line and the semi-classical collapse leading to a regular black hole developed in \cite{Malafarina:2022oka} is shown as the dashed line. The parameters are fixed for illustrative purposes to $m_0=1$ and $\xi=0.01$.} \label{a}
\end{figure}

We now consider the scenario of dust collapse, therefore taking $p=0$, $\epsilon \propto a^{-3}$, and $\rho = \epsilon$. In accordance with the above discussion we interpret the variable $d$ as the proper distance as in \cite{Bonanno:2000ep} so that
\begin{equation}
d\sim \frac{r^{3/2}}{\sqrt{m}}\sim\frac{1}{\sqrt{\epsilon}},
\end{equation}
where $r$  is the radial distance. At last we thus obtain the following expression for the behavior of $G(\epsilon)$:
\begin{equation}
\label{G}
G(\epsilon)=\frac{G_N}{1+\xi \epsilon},
\end{equation}
where we introduce the dimensionful scale $\xi$, and we include the pure number $g_\ast$ in the definition of $\xi$. It is important to note that, in general, we would expect $\xi \sim 1/m_{\rm pl}^4$, but currently, there is no clear method to determine $\xi$ from first principles, and this parameter should be constrained from observations. 
Setting $ 8 \pi G_N=1$, we obtain: 
\begin{equation}
\chi(\epsilon)=\frac{\log(1+\xi \epsilon)}{\xi\epsilon}, \quad \Lambda(\epsilon)=\frac{\log(1+\xi \epsilon)}{\xi}-\frac{\epsilon}{1+\xi\epsilon}.
\label{runa}
\end{equation}
Importantly, in the classical limit, achieved for $\xi\rightarrow 0$, we recover $\chi=1$ and $\Lambda=0$, as expected. Figure \ref{fig:V} illustrates the potential $V(a)$ for the running coupling defined in equation \eqref{G}, as compared to the Oppenheimer-Snyder-Datt (OSD) case.
Reducing our formalism to the OSD model \cite{1939PhRv...56..455O, 1938ZPhy..108..314D}, which describes non-interacting dust particles with $p=0$, we have $T_{\mu\nu}=\epsilon u_\mu u_\nu$, $\epsilon = 3 m_0 / a^3$
and $m_{\rm eff}\rightarrow m_0$.
Equation \eqref{V(a)} yields $V(a)=-m_0/a$  
recovering the usual equation for the scale factor of homogeneous dust:
\begin{equation}
\dot{a}=-\sqrt{\frac{m_0}{a}-K}.
\end{equation}
It is worth noting that bound collapse is obtained for $K>0$, while the marginally bound case has $K=0$. Interestingly, to achieve singularity resolution at the end of the collapse, the usual energy conditions must be violated. In many models available in the literature, the growth of negative effective pressures leads to repulsive effects that halt the collapse. However, in our case, such a term is due to the running cosmological constant, which generates repulsive effects, allowing the collapse to come to a halt. For $\xi\neq0$ and \eqref{G}, we obtain:
\begin{equation}
\label{eom}
\dot{a}=-\sqrt{\frac{\log(1+3m_0\xi/a^3)}{3 \xi} a^2-K}.
\end{equation}
The solution to this equation for marginally bound collapse ($K=0$) is shown in Figure \ref{a} and compared with the OSD model and the semi-classical collapse model developed in \cite{Malafarina:2022oka}. At large times, the scale factor behaves as:
\begin{equation}
a(t) \sim e^{-t^2/4 \xi}, \quad t\rightarrow \infty,
\end{equation}
indicating that $a=0$ is never reached at any finite time, and the spacetime is geodesically complete. As $t$ approaches infinity, the scale factor tends to diminish exponentially, 
ensuring the avoidance of singularities within  any finite time frame.

To implement the matching of the collapsing matter cloud described above with a suitable exterior, we employ the formalism developed by Israel in \cite{1966NCimB..44....1I}, which was further refined by Senovilla and others in \cite{1992PhRvD..45.2732F, 1996PhRvD..54.4862F}.
We consider the matching across a comoving boundary $r=r_b$ in the interior, which corresponds to a collapsing boundary $C_b(t)=C(t,r_b)=r_ba(t)$. 
The induced metric on the matching surface $\Sigma$ in comoving coordinates can be expressed as:
\begin{equation}
ds^2_\Sigma=-dt^2+r_b^2a^2d\Omega^2.
\end{equation}
For the exterior, we consider a generic static and spherically symmetric line element in $\{T,R,\theta,\phi\}$ coordinates  written as:
\begin{equation}
ds^2 = -f(R) dT^2 +\frac{1}{f(R)} dR^2 +R^2 d\Omega^2,
\end{equation}
where $f=1- 2 M(R)/R$, and we assume a continuous matching between the two geometries. The continuity condition uniquely determines the form of $M(R)$ in the exterior. Specifically, if the collapsing boundary is parametrized by $R=R_b(T)$, the induced metric on the boundary becomes:
\begin{equation}
ds^2_\Sigma =
-\left[f(R_b)-f(R_b)^{-1}\left(\frac{dR_b}{dT}\right)^2\right]dT^2+R_b^2d\Omega^2.
\end{equation}
The matching conditions for the metric functions on the boundary surface $\Sigma$ immediately provide the relation between $t$ and $T$ on $\Sigma$  and the condition $R_b(T(t))=r_ba(t)$.
The second fundamental form for the interior metric in comoving coordinates is 
\begin{equation}
K_{tt}^-= 0, \quad
K_{\theta\theta}^-= r_ba\sqrt{1-Kr_b^2}.
\end{equation} 
From the extrinsic curvature on the exterior we  obtain
\begin{equation}
K_{tt}^+= -\frac{1}{2}\frac{2\ddot{R}_b+f_{,R}(R_b)}{\Delta(R_b)}, \quad 
K_{\theta\theta}^+= R_b\Delta(R_b),
\end{equation}  
with $\Delta(R_b)=\sqrt{1-2M(R_b)/R_b+\dot{R}_b^2}$,
so that on imposing 
\begin{equation}
\label{match1}
[K_{tt}]=K_{tt}^+-K_{tt}^-=0, \quad
\left[K_{\theta\theta}\right]= K_{\theta\theta}^+-K_{\theta\theta}^-=0,
\end{equation}
the functional form of $M(R)$ can be obtained. 
Finally, we arrive at the most important result of our investigation:
\begin{equation}
M(R) = \frac{R^3}{6 \xi} \log \left(1 + \frac{6 M_0 \xi }{R^3}\right),
\end{equation} 
where the matching implies $m_0 r_b^3 = 2 M_0$. 
This expression describes the Misner-Sharp mass pertaining to the static exterior, originating from an interior undergoing gravitational collapse. The dynamics of this interior are guided by an effective Lagrangian that incorporates the Asymptotically Safe (AS) nature of gravitational interaction at Planckian energy scales.

Importantly, the classical limit is recovered for $\xi \rightarrow 0$ (or equivalently for $R\rightarrow \infty$), leading to:
\begin{equation}
M(R)=M_0-\frac{3 M_0^2 \xi}{R^3}+\frac{12 M_0^3\xi^2}{R^6}+O(\xi^3),
\end{equation}
as expected. In the low-energy limit, the Schwarzschild solution is regained. Notably, in the small $R$ regime, $M(R)$ behaves like
\begin{equation}
M(R)=    
\frac{1}{6\xi} R^3 \log \left(\frac{6 M_0 \xi }{R^3}\right)+\frac{R^6}{36 M_0 \xi ^2}+O\left(R^7\right)
\end{equation}
moreover, as $R\geq R_b= r_b a(t) $ and $a(t)>0$ always, our solution remains everywhere regular, avoiding any singularities. 
This result is of significant importance as it demonstrates the compatibility of the collapsing matter interior with our AS effective Lagrangian model in producing a regular black hole exterior.
 \begin{figure}[tt]
\includegraphics[width=0.97\linewidth]{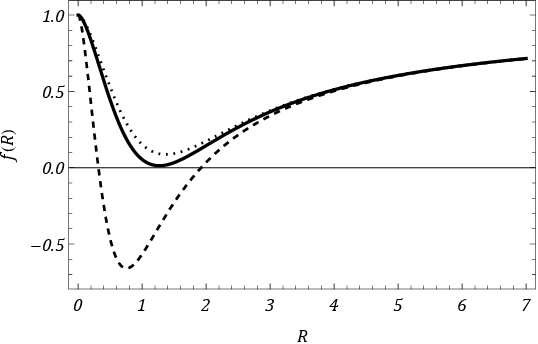}
\caption{The behavior the metric function $f(R)$ for different values
of the parameter $\xi$ for $M_0=1$. For $\xi< \xi_{\rm cr}$ there are an inner and an outer horizon, corresponding to the two solutions of $f(R)=0$ (dashed curve for $\xi=0.1$) and for $\xi= \xi_{\rm cr}\approx 0.45$ there is only one horizon (thick curve). For $\xi> \xi_{\rm cr}$  the two horizons disappear and one is left with a scalar remnant (dotted curve for $\xi=0.6$). } \label{fig:fr}
\end{figure}

The horizon's position is determined by solving the transcendental equation $f(R)=0$, which can yield intriguing results. Notably, for any specified value of $M_0$, a critical threshold $\xi_{\rm cr}$ 
exists. When $\xi <\xi_{\rm cr}$, an event horizon becomes evident, accompanied by an inner horizon at smaller values of the radial coordinate. However, for $\xi >\xi_{\rm cr}$, a scalar remnant emerges, as illustrated in Figure \ref{fig:fr} for $M_0=1$.

\begin{figure}[tt]
\includegraphics[width=0.97\linewidth]{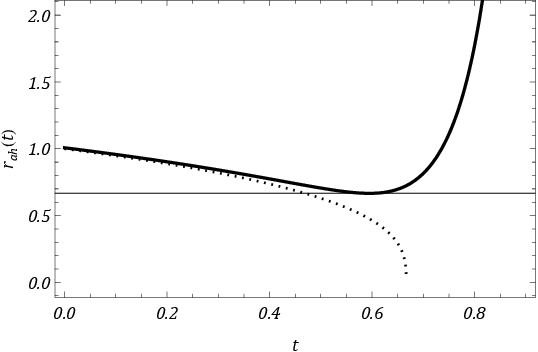}
\caption{The thick line shows the apparent horizon $r_{\rm ah}(t)$ for marginally bound dust collapse in the Asymptotically Safe model with running $G$, the dotted line shows $r_{\rm ah}(t)$ for the OSD model with $8\pi G_N=1$. Notice that in the AS case the apparent horizon reaches a minimum value $r_{\rm min}$, as shown by the thin horizontal line, and then grows. This means that for a boundary $r_b={\rm const}.$ taken at $r_b<r_{\rm min}$ the whole collapse process is not covered by any horizon. The parameters are fixed for illustrative purposes to $m_0=1$ and $\xi=0.01$.} \label{fig:rah}
\end{figure}

We expect for the global solution, that the evolution of the causal structure in the interior matches the formation of the causal structure for the exterior geometry. The condition for the formation of trapped surfaces in the interior may be obtained from
\be 
1-\frac{F_{\rm eff}}{C}=1-r^2(\dot{a}^2+K)=0,
\ee 
which implicitly gives the curve $r_{\rm ah}(t)$ describing the comoving time $t$ at which the shell $r$ becomes trapped:
\be 
r_{\rm ah}(t)=\frac{1}{\sqrt{\dot{a}^2+K}}.
\ee 
It is clear that in the OSD case as $\dot{a}$ diverges the apparent horizon curve tends to zero. On the other hand in the AS case we have that $\dot{a}$ goes to zero asymptotically and therefore $r_{\rm ah}\rightarrow 1/\sqrt{K}$.
In the marginally bound case, for any given $\xi$ 
we have that $r_{\rm ah}$ reaches a minimum value $r_{\rm min}$ and then grows to infinity. Then there are values of the boundary $r_b>r_{\rm min}$ that lead to $r_{ah}$ crossing the boundary twice thus creating two horizons in the exterior. Accordingly there is a critical value $r_b=r_{\rm min}$ for which only one horizon exists and for values smaller than $r_{\rm min}$ no horizon forms throughout collapse. The apparent horizon curve is shown in Figure \ref{fig:rah}.

Our solution represents a significant alternative to present models of regular black holes. It is built upon the assumption that 
black hole solutions observed in Nature are sourced by a matter interior whose evolution is non-singular due to the antiscreening of the gravitational constant at small distances \cite{Polyakov:1993tp}, according to a specific renormalization group trajectory terminating at the Reuter fixed point of AS gravity. This mechanism is implemented using an effective Lagrangian that incorporates a multiplicative coupling with the matter component. Although in this work, we considered an idealized model of matter consisting of a pressureless fluid, our framework can be generalized to incorporate more realistic equations of state 
and more accurate RG trajectories, providing a consistent description of the matter component. We intend to address these issues in future investigations.

\section*{Acknowledgement}
DM would like to thank Catania Astrophysical Observatory - INAF -  for warm hospitality during the preparation of the manuscript.
DM acknowledges support from Nazarbayev University Faculty Development Competitive Research Grant No. 11022021FD2926.

\end{document}